# Secure SQL Server – Enabling Secure Access to Remote Relational Data

The *Secure SQL Server* – SecSS, is a technology primarily developed to enable self-service governance of states, as described in (Paulin 2012). Self-service governance is a novel model of governance that rejects service-based public administration and instead proposes that governed subjects manage their legal relations in a self-service manner, based on ad-hoc determination of eligibilities. In this article we describe the prototype SecSS and its evaluation in a complex governmental scenario.

## 1   Introduction

Our general research follows the ideological goal of designing a society that does not need human-managed bureaucracy for governance. In (Paulin 2012) we argue that governing states and other forms of human societies can be managed by the citizens/subjects themselves, with no need for bureaucratic intervention. In order to achieve self-service governance (ss-Gov), it is essential to focus on the cornerstones of governed societies, whereby the essential information needed for governance is – besides the identities of the governed subjects, knowledge about legal relations between members of the particular society. Provided that such information is available in form of discreet information within distributed information systems, individual eligibilities to do or obtain something in a given context can be determined algorithmically, which consequently eliminates the need for manual evaluation of given eligibilities.

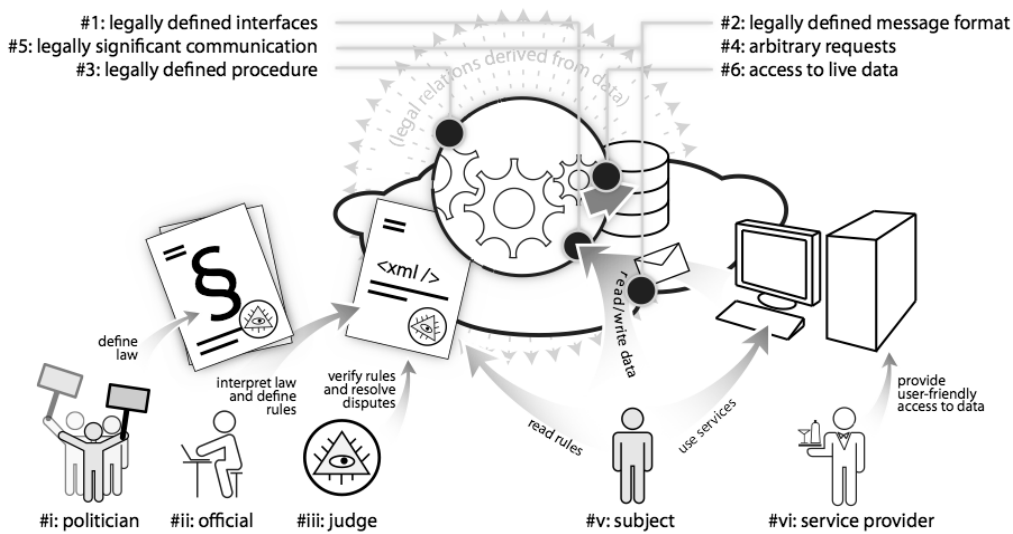

**Figure 1: The principles and stakeholders of ss-Gov**

In ss-Gov, data required for governing must be stored in databases, to which subjects have read and write access. Trough self-service manipulation of the stored data, subjects themselves manage their legal relations and rights, thus for example somebody can gain the right to drive cars or the eligibility to receive child support (cf. the examples in Paulin 2012). Eligibilities in ss-Gov are not stored, but calculated ad-hoc and are based on live data – the later being crucial for the existence of eligibilities.

Distributed databases ("registries") and the interaction with them are essential for ss-Gov – not only from the organizational, but also from the legal perspective. In order to comply with basic legal principles, most notably the principle of legality, six requirements have been defined that any ss-Gov system must fulfill (Paulin 2012, 2.2.1): The system must legally define (1) the interface of the system on the Internet – e.g. URI, (2) the formats of the exchanged messages and (3) the procedure how messages will be handled. Further such system (4) must accept arbitrary requests in a standardized artificial language, (5) must guarantee legally significant communication and (6) must allow direct access to core data restricted only by legal requirements.

Furthermore, ss-Gov knows six stakeholders (Paulin 2012, 2.2.3), whose needs must be satisfied: (i) politicians, who make descriptive legislation, (ii) officials, who translate legislation into a system of databases, data and rules of access to this data, (iii) judges, who resolve disputes in case that legislation was not correctly translated by the officials, (iv) administrators who maintain the integrity of the ss-Gov systems, (v) subjects, who access the data and (vi) service providers, who offer products – e.g. graphic user interfaces, programming libraries or notary services, to facilitate the manipulation with ss-Gov data for the subjects.

In the present article we aim to describe how such a registry could be realized using currently available technology, whereby our utmost priority shall be to fulfill the six technical requirements and satisfy the six stakeholder types stated above. We shall therefore find suitable technologies for storing and accessing data as well as for optimal enforcement of access-restrictions. In order to prove the feasibility of the chosen technologies, we shall present a prototype ss-Gov system.

We chose the design science methodology (March and Smith 1995; Hevner et al. 2004) for the presented research, considering the guidelines by Roel Wieringa (2009). Wieringa puts emphasis on the *regulative cycle* as a framework for the logic of practical problem solving, whereby this cycle consists of four phases: problem investigation, solution design, design validation and the optional solution implementation.

The research presented in this paper follows the goal to construct the concrete ss-Gov prototype, which is the first system instantiating ss-Gov and thus out of competition (the feasibility of ss-Gov as such has been demonstrated in (Paulin 2012)). According to Wieringa (2009), a design solution/artifact is successfully validated, if it meets the criterion of *internal validity*, i.e. if the effect of the solution would satisfy stakeholder criteria, which are to be elaborated trough the problem definition. In our case the solution must meet both the requirements of the six stakeholder roles, as well as the six requirements for ss-Gov registries summarized hereinabove. Designing an adequate solution is further bound to several important technology-related *knowledge questions* (cf. Wieringa 2009, 2), which we have to resolve beforehand, such as which technologies to use for identifying subjects, for querying and communicating data, and how to define and apply restrictions of access.

We further define stakeholder needs and discuss the technological alternatives in chapter 2 and justify our decisions. In chapter 3 we describe the designed solution – the Secure SQL Server and evaluate it finally in chapter 4 using a scenario of governing children playing in a sandbox.

## 2   Stakeholder needs & technologies

SS-Gov defines six stakeholder roles, as illustrated in Figure 1, however only two of them are relevant for designing the ss-Gov registry, namely the officials and the subjects. Politicians are not relevant for the design, because they only generate rules that must be translated by the officials into digital form. It is the later that must deal with the standards and semantics of the technical solution and it is also the officials that take responsibility for their translation. Also the role of administrators is not important from the design point of view, as their only task is to provide the integrity of the system, without interfering with the data.

Service providers do not play an active role, however their existence implies the requirement for proxy access to data, i.e. access on behalf of subjects. Thus for example, a notary acting as a service provider could commit a real-estate transaction on behalf of her client, whereby the respective ss-Gov registry would receive the request for data manipulation from the notary instead of the owner of the real estate. On the other hand, power given to somebody for representation either explicitly or by law (e.g. parents representing children) is only information that must be determinable by ss-Gov trough dedicated registries and regulated by law. Hence, also the role of service providers does not impose any requirements on the design that would not be already expressed by the six technical & legal requirements.

Subjects use ss-Gov to manipulate data based upon which concrete eligibilities in a given context are determined. Any request received by an ss-Gov registry must therefore be a non-repudiable expression of will of its sender and the sender's identity must be revealed to the registry. This implies the requirement of using an electronic ID (e-ID) and electronic signature to assure non-repudiation of origin in any communication with ss-Gov.

Two questions need to be answered so far: which technology shall we use to identify users and assure non-repudiable messages? And: which technology shall be used to store the data so that an arbitrary read-write access restrained only by electronic rules can be provided? In the next two subchapters, we shall find first the answer to the later question and then to the former.

### 2.1   Managing read-write access to data

Relational databases (Codd 1970) are the dominant structure for storing data inside information systems today (cf. Bain 2009). From the perspective of the present research it is of no importance how the data is represented inside the databases; what is important however is that data from such databases is usually accessed trough relational database management systems (RDBMS) using the standardized query language SQL (Chamberlin and Boyce 1974; ISO/IEC 9075-1 2008).

Trough SQL, data structures are defined, created, manipulated and queried using relational algebra; SQL can be used to read and write data, as well as to perform complex queries to filter and combine sets of data, etc. Because the relational data model does not imply how data is stored physically in the system, SQL remains platform and vendor independent.

There are no explicit means to regulate access to data provided by SQL as such. Vendors of RDBMSs implement their own solutions, which often result in the management of access privileges for various users, whereby such privileges are restricted only to the level of schemas ("databases") and relations ("tables")[1].

More detailed rules for managing the access to data are usually hardcoded in the business logic of the front-end applications. However, as argued in (Paulin 2012, 1.2), hardcoded rules are no sustainable solution for governing and must therefore be avoided if possible.

This requirement factually means that a way must be found how fixed business logic can apply frequently changing and not-known-in-advance rules to not-known-in-advance request. Provided that requests are in SQL, it is possible to express access restrictions in form of subsets of data to which the requester would have unlimited access. SQL allows calling a query against sub-queries, whereby first each sub-query is executed, resulting in a virtual relation; the outer query is then executed against such virtual relation. As this technique was applied also for the SecSS, it will be described in more detail in chapter 3.

SQL would therefore be a feasible option that would satisfy both requirement #4 (arbitrary requests in standardized language) and #6 (access to core data limited only by law).

An alternative technology is the *Resource Description Framework* (RDF), which plays an important role in the Linked/Open Data movements (cf. Veljković, Bogdanović-Dinić, and Stoimenov 2011) and the Semantic Web (Berners-Lee et al. 2001; cf. Vitvar, Peristeras, and Tarabanis 2010). RDF is a syntax-neutral framework for storing data in subject-predicate-object *triples*, whereby the *subject* denotes the described resource, the *predicate* its attribute and the *object* its value. Various syntaxes for expressing and communicating RDF exist, such as RDF+XML, N3 (Berners-Lee et al. 2008) or Turtle (Beckett and Berners-Lee 2008).

Berners-Lee (2009) sees RDF as an essential technology for publishing data on the Web in a way that would allow intelligent reuse and integration of remote sources of data, which would result in the Semantic web. However, as Patel-Schneider (2010) points out, RDF was designed to contain structure, not semantics; the later being introduced around 2004. (The semantic aspects of RDF are for ss-Gov not relevant. The semantics of data stored in ss-Gov are defined and determined by law, thus there is no need for meta-data representing semantics or such kind.)

---

[1] Cf. the user management of MySQL, MSSQL, Oracle, etc.

RDF is a model that bases on the graph theory (cf. Hayes and Gutierrez 2004; cf. Berners-Lee 1998) and is thus in sharp contrast to the relational model. Although RDF was not designed as a database technology, but only as a model to represent data, we might consider the use of triple-stores, i.e. dedicated graph databases for storing RDF. Graph databases are not new and they are known for achieving better performance than relational databases and therefore are used for high load scenarios by global enterprises like Google, Facebook, LinkedIn or Amazon (Vicknair et al. 2010).

Graph databases in comparison to relational databases lack a common query language. Thus, Angles & Gutierrez (2005) list a wide assortment of languages used to retrieve data from graph databases: G+, GraphLog, Gram, PaMal, GraphDB, Lorel, plus rivaling languages dedicated for querying RDF: RQL, SquishQL, RDQL, RDFQL, TRIPLE, Versa, SeRQL, RXPath.

In 2004 the SPARQL query language for RDF appeared, which since 2008 is a W3C recommendation. Although originally only a language for read-querying data, SPARQL later received the amendment SPARQL-Update, which allows write access to RDF data (Seaborne et al. 2008).

Both SQL and SPARQL have similar characteristics and allow read/write access to data, as well as sub-queries. While the former has been an established technology for several decades, the later was designed with the distributed architecture of the Web in mind. However, while in theory both options seem equally feasible, we chose SQL, as it is a mature, standardized technology with undisputed market dominance.

### 2.2   Identification and authorization

For each message (both request and reply) transmitted in ss-Gov it is essential that the receiver can be sure that the message is non-repudiable; further, the rules to be applied must be aware of the requester's identity in order to provide a personalized authorization of access – it is therefore essential that the identity is a non-repudiable part of each message.

We considered a number of technologies used for identification and authorization: first to mention are various single-sign-on (SSO) technologies that allow users to identify themselves to heterogeneous servers using a single identity. David *et al.* (2011) list Kerberos, OpenID and Snap2Pass as notable options for SSO.

Kerberos (S. P. Miller et al. 1987) is one of the first SSO protocols for authorizing access to remote services in a network. Its core architecture uses two tightly synchronized servers, the authentication server that authenticates the identity of the client, and the "ticket granting server" that issues time-limited tokens ("tickets") for accessing services. David *et al.* (2011) note that the protocol is not only highly complex but from today's perspective also not adequately secure any more.

OpenID (cf. Recordon and Reed 2006) is an open standard for SSO over the Web. It does not depend on a central authority that would issue identities, thus users can choose from various identity-providers. The individual authentication towards the identity-providers can require the user to enter a basic username/password combination, but also advanced biometric and smartcard authentication are possible. After authenticating the user, the server issues a Yadis (J. Miller 2006) document, which is then used for identification towards applications. OpenID is significant because of its popularity and it is used by Google, Yahoo, MySpace, AOL, PayPal, VeriSign, etc. However, OpenID is neither designed to provide message integrity nor the non-repudiation of its content, thus ss-Gov requirement #5 is not met.

Also Snap2Pass is not suitable due to the same reason. Snap2Pass (Dodson et al. 2010) is a protocol that authenticates the user by presenting her a randomly generated QR code which the user must capture by her mobile device on which a shared secret is stored. Using that shared secret and the unique token received by capturing the QR code a hash-based message authentication code (HMAC) is generated and returned to the server.

In another survey Akram & Hoffmann (2008) consider also the aspect of non-repudiation when evaluating six popular technologies – SAML, OpenID, CardSpace, Shiboleth, Higgins and Liberty, however they find only Higgins capable to adequately provide this feature. Higgins was also the best recommendation for user-centric identity management in a survey conducted by Maliki & Seigneur (2007).

The broad assortment of available and constantly evolving technologies for identity management (cf. Lampropoulos and Denazis 2011) motivates us to chose conservatively rather than progressively. We chose the well-established X.509 ITU standard and the PKI infrastructure for identity management.

The X.509 standard provides the concept of digital certificates that are issued by the identity provider, which should be a trusted third party (TTP). Each certificate is digitally signed by its issuer and usually contains a public key, which corresponds to a private key known only by the owner of the certificate, thus enabling the owner to digitally sign and encrypt communication. Under certain conditions, digital signatures are non-repudiable and enjoy the same power as their traditional, handwritten counterparts (cf. Blythe 2005).

## 3   SecSS

The Secure SQL Server (SecSS) prototype is a Web application built using Microsoft .NET 4.0 WCF technology (cf. Mackey 2010). SecSS acts as a neutral agent that receives requests for querying data, expressed as digitally signed SQL queries. After receiving a query, SecSS verifies its integrity and extracts the identity of the signer; it then neutrally applies rules to it as sub-queries, and finally executes the full query against the database. The RDBMS of our choice for the SecSS prototype is MySQL 5.1.

Complementary to the SecSS server we developed a user-friendly Ajax application that runs in contemporary Firefox web browsers. We chose this browser after carefully evaluating the popular alternatives because it offers cross-platform performance according to latest Web standards and further provides built-in (proprietary) cryptographic functionality – the *crypto* object (Anon. 2005), which can be consumed trough JavaScript to e.g. digitally sign text using IETF's *Cryptographic Message Syntax* (CMS) standard; comparable cross-platform web browsers like Safari or Chrome, which both base on the Apple Webkit engine, do not provide such functionality.

The Ajax client features two separate user interfaces – a graphical user interface (GUI) for point-and-click/drag-and-drop interaction and a text-only interface for submitting arbitrarily generated queries.

### 3.1   Handling incoming requests

The SecSS expects requests packed in a custom JSON structure, as shown in Table 1. We choose between the well-known standards XML and JSON for the format of messages exchanged between the client and the server and decided on the later, as the client application can easier generate and consume this format.

```
{
    SQL:        /* content of the request expressed in SQL */,
    Pkcs7:      /* Base64 encoded Pkcs7 signature of the SQL */,
    Comment:    /* informative description of the request */
}
```

**Table 1: JSON structure of the digital request**

The request object provides three fields: "SQL" contains the plaintext SQL query as either generated trough the graphical user interface of the client, or designed by the user herself; the "Pkcs7" field contains the Base64 encoded digital signature of the SQL query, and the "Comment" field may contain additional natural language comments, which however are just informative.

The subject making the request is free to make whatever SQL query she likes. Such query may consist of many SQL statements, which could be even malicious SQL injection attempts (cf. Anley 2002) intended to harm the system.

Each received request is first checked for its integrity and validity of the digital signature, from which we also excerpt the digital certificate of the requester. Based on the digital certificate a SHA-256 hash is computed and encoded in Base64, which uniquely identifies the subject internally. This unique identifier allows us to use the identity for ad-hoc personalization of access rules.

To analyze the received SQL we use the *Gudu Software SQL Parser*, a competent grammar parser, which allows us to separate individual statements, validate the received SQL and extract information about which types of queries (e.g. SELECT,

INSERT, UPDATE,...) are requested and which relations and attributes are being accessed.

The SecSS prototype implements only four frequently used statement types: SHOW, SELECT, INSERT and UPDATE. The SHOW query, which allows subjects to get basic information about the structure of the relations, is executed with no restrictions. The other three types are queries to which rules are applied before execution in form of sub-queries, as described below.

The received request is handled immediately and also the response is generated fully automated. Responses issued by SecSS are transmitted as JSON objects as shown and explained in Table 4.

---

Original request (putting a child with a toy into the sandbox):

```
INSERT INTO sandbox (name, toy) VALUES ('Loys', 'ball');
```

Rule (a child may play only with toys suitable to its age):
*Remark: the rule imposes that the attributes* `toy` *and* `name` *are assigned trough variables* `@toy` *and* `@name`.

```
WHERE @toy IN
(SELECT t.toy FROM toys t
      WHERE t.ageLimit < (SELECT c.age FROM children c
            WHERE c.name = @name))
```

Transformed request:

```
SET @name = 'Loys';
SET @toy  = 'ball';

INSERT INTO sandbox (name, toy)
SELECT @name, @toy FROM DUAL
      WHERE @toy IN
      (SELECT t.toy FROM toys t
            WHERE t.ageLimit < (SELECT c.age FROM children c
                  WHERE c.name = @name));
```

**Table 2: Transformation of the INSERT statement according to stored rule.**

---

### SELECT

When a SELECT query is requested, then the SecSS prototype verifies whether the subject has explicit permission to read each of the requested attributes. In case that all attributes (*) are requested, permission to access all attributes must be explicitly granted. The request is denied if for at least one of the requested attributes no read-access has been granted.

Usually, access to attributes will be granted conditionally. Thus for example, information about employees of a state organization may be given only for full-time employees. Such condition will present a rule expressed as a sub-query and will be applied whenever read access to any field of the relation containing the requested data is performed.

If the request is conditional, thus if it contains a WHERE clause, then such clause is isolated using parentheses for security reasons.

### INSERT

The same as when read access is requested, we first check at the INSERT statement if write access to the particular attributes has been granted. We then extract each attribute to which a value is being assigned and define it as a variable.

The INSERT statement is fully reconstructed. If the rules impose filtering, then the request is executed against a virtual (DUAL) relation, as shown in Table 2.

### UPDATE

Also in UPDATE queries SecSS first checks if apropriate permissions for the requester exist. The logic for transforming statements is similar to the one applied for INSERT statements. Table 3 shows how an UPDATE request is transformed according to the rules.

Eventually it may happen that values in the UPDATE statement are themselves results of nested queries instead of fixed values. Table 5 shows how SecSS handles such situation.

---

Original request (giving a child another toy):

```
UPDATE sandbox SET toy = 'squirrel'
      WHERE name = 'Loys';
```

Rules (a child may play only with a (1) toy suiting its age, (2) which is not occupied):
*Remark: the first rule requires attributes* `toy` *and* `name` *to be assigned trough variables* `@toy` *and* `@name`; *the second rule requires the existence of variable* `@toy`, *which is the same as the previous.*

```
WHERE @toy IN
(SELECT t.toy FROM toys t
      WHERE t.ageLimit < (SELECT c.age FROM children c
            WHERE c.name = @name))

WHERE @toy NOT IN
(SELECT s.toy FROM sandbox s)
```

Result of transformation:

```
SET @name = 'Loys';
SET @toy  = 'squirrel';

UPDATE sandbox SET toy = @toy WHERE (name = @name)
AND @toy IN
      (SELECT t.toy FROM toys t
            WHERE t.ageLimit < (SELECT c.age FROM children c
                  WHERE c.name = @name));
AND @toy NOT IN
      (SELECT s.toy FROM sandbox s)
```

**Table 3: Transforming the UPDATE query.**

---

### 3.1   The Electronic Legal Act

Rules are stored in *Electronic Legal Acts* (ELAs), a special XML-based document format developed for the requirements of the SecSS. The ELAs are publicly readable documents, issued and signed by the responsible public official, which contain rules of access to data stored on the particular server. The publicness of the ELAs is essential as it allows everybody to examine the rules that apply to her and act accordingly.

The XML schema of ELAs is shown in Figure 4 and Figure 5. The root element (*Configuration*) provides the elements *Connection, Restrictions* and *Permissions*. The *Connection* element is provided to hold the full connection string to accessing the database, however it is not used in the prototype.

Restrictions for accessing individual attributes (hereinafter: fields) are stored inside the element *Restrictions* as a collection of *Restriction* elements (cf. Figure 4). Each restriction has its unique ID (attribute *@Id*), which is used to address the restrictions from inside rules that regulate access to individual fields. The attribute *@type* defines at which request the restriction should be applied; the prototype supports two different types of request: *SELECT* and *INSERT/UPDATE*. Thus, a restriction of type "SELECT" shall be applied only when read access is requested.

```
{
    Results:            /* collection of resulting relations */
    [{
        ExecutedSQL:      /* transformed statement */,
        RequestedSQL:     /* original SQL statement */,
        Rows:             /* collection of returned rows */
        [{
            Name:         /* attribute name */,
            Value:        /* attribute value */
        }]
    }]
    Feedback:           /* auto-generated feedback */,
    GenerationDate:     /* date and time of the response */,
    OK:                 /* success? (true/false) */
}
```

**Table 4: JSON structure of the response**

Restrictions are applied in form of sub-queries, as described in the previous chapter. The sub-query defines a subset of data, a virtual relation, over which the requested query is executed. The attributes *@field* and *@use* define how the sub-query should be integrated into the final query: *@field* defines, which value shall be searched-for in the result of the sub-query and *@use* determines the method for search, whereby "*IN*" and "*NOT IN*" are supported. Table 2 shows how an incoming request is transformed according to rule *#suitableAge* from Figure 4.

Rules may require that certain fields or variables be addressed in the request. In such case (cf. Figure 4) rules contain elements *var* in which the variable names and addressed fields are defined. In the example from Table 2 the fields *name* and *toy* have been assigned the values "Loys" and "ball" and both values are "captured" by the provided variables *@name* and *@toy*, which are later addressed from within the sub-query, e.g.: *WHERE c.name = @name*. These required variables enable rules to be context-aware.

The element *Permissions* hierarchically structures information to what fields access-permissions have been defined. The ELA generally regulates access for the entire endpoint (one RDBMS), which may contain multiple schemas. *Permissions* therefore supports many *Schema* elements, each *Schema* can contain many *Tables* and each *Table* many *Fields*. Access permissions have to be explicitly regulated on the level of individual fields – the rationale of this design is to force ELA designers to think about every single field to which they grant access.

---

Original request (child with the identifier 1008984500000 receives toy #15):

```
UPDATE sandbox s LEFT JOIN children c ON s.name = c.name
SET s.toy =
    (SELECT t.toy FROM toychest t WHERE t.id = '15')
WHERE c.emšo = '1008984500000';
```

Transformation result:
*Remark: the rules impose the variables* `@toy` *and* `@name`*, which provide values for the attributes* `sandbox.toy` *and* `sandbox.name`*; the requester however provides only a value for* `sandbox.toy`*.*

```
SET @name = (SELECT c.name FROM children c
                WHERE c.emšo = '1008984500000');
SET @toy  = (SELECT t.toy FROM toychest t WHERE t.id = '15');

UPDATE sandbox s LEFT JOIN children c ON s.name = c.name
SET s.toy = @toy
WHERE (c.emšo = '1008984500000')
AND @toy IN
    (SELECT t.toy FROM toys t
        WHERE t.ageLimit < (SELECT c.age FROM children c
            WHERE c.name = @name));
AND @toy NOT IN
    (SELECT s.toy FROM sandbox s)
```

**Table 5: Resolving undefined variables in UPDATE statements**

---

Each field can contain multiple permissions (*Permission*) and each permission has the mandatory attribute *@user*, which defines to which requester the containing rules have to be applied. To regulate permissions for public access, the value of *@user* is "anon".

In the prototype, user management is static, i.e. the users to which special regulations apply are explicitly defined in the ELA. However, a dynamic solution would be imaginable, where the identity identifier would be the result of a prior SQL query – in such case restricting access to a position (e.g. prime minister) or membership to a group (e.g. police force, faculty staff, voting body) would be imaginable. A further not yet realized option would be to implement a "global" SQL variable *@sx_identity* that would expose the identity of the requester to the sub-queries containing the rules; better context-awareness could be achieved by exposing further global variables providing e.g. the IP address.

Particular permissions can be restricted trough rules, which SecSS applies in form of sub-queries. These restrictions are applied by defining a series of *Apply-Restriction* elements, which refer to the particular rules using the *@ref* element referring to the *@Id* of the rule. The defined restrictions are applied in their numerical order.

### 3.2   Summary: effective personalized regulation trough sub-queries

The main focus of SecSS is on sub-queries trough which read/write access to data needed for governance is regulated. The dynamic sub-queries as used by SecSS represent context-aware rules that are applied cumulatively to an incoming request, thus presenting the requester a set of virtual relations of data to which she has the desired access. In this way it is possible to create a dynamic set of rules, which adopt themselves to the situation on run-time.

Personalized regulation of access to data in modern praxis is usually done trough workflow-based business logic, which has been defined at design-time of the application; this way a very fine-grained and performance-effective regulation mechanism can be designed, however any change of rules or underlying data structure requires a redesign of the business logic itself, which might be questionable from the legal perspective (cf. Paulin 2012).

Another way to personalize access to data in relational databases has been patented by Garrison (1998), who describes a method to regulate access to individual fields/rows. His concept requires a "security data table" that would contain particular information about which user has access to which fields. In our opinion however, this approach would result in an exorbitantly large metadata table with questionable added value.

The approach implemented by SecSS relies on relational algebra. An important strength of the described method lies in the opportunity to regulate personalized access for a-priory unknown users without storing any information about them. Neither the identity of the user, nor the exact subset of data to which the user requests read or write access can be predicted in advance, nonetheless SecSS is able to effectively regulate in accordance to given rules.

Absolute flexibility in regulation satisfies the ss-Gov requirements #4 (arbitrary requests in standardized language) and #6 (access to core data limited only by law). SecSS further satisfies the ss-Gov requirement #2, which demands that the format of exchanged messages is to be legally defined, as well as requirement #3 that demands a legally definable procedure how messages will be handled; the later requirement is achieved by separating the definition of rules from the business logic, which' code therefore can be politically approved and published in the same way as legislation and remain unchanged for generations.

## 4   Proof-of-concept: the sandbox

In order to evaluate the SecSS, we chose the scenario of a sandbox – both a popular term from computing as well as a complex world of legal relations in the reality. The sandbox from our scenario is centered inside a playground and presents a geographically controllable area of screen pixels inside which virtual children can play with their virtual toys.

The playground is a schema containing three relations: *children, sandbox* and *toychest*, as shown in Figure 2. This virtual habitat resembles a real society – it contains subjects (the children), which inhabit a territory (the sandbox) and have relations to objects (the toys). Children can be placed in and removed from the sandbox and they can receive/give back toys into the toy-chest. All relations among the subjects and objects in the playground are defined as data in the database and every change of them is stored.

### *The rules*

As it is in any real society, also the playground knows certain rules that regulate the interaction. The rules that apply to the playground protect the interests of the children, namely their right to data privacy and possession of goods. The rules are as follows:

1. Everybody can *read* any data with exception of the dates of birth of the children, which are protected personal data.
2. A toy that is already in use must not be given to any other child.
3. Everybody may give any child a toy to play with; however the children can receive only toys for which they are old enough.
4. Everybody may place a new toy into the toy-chest.
5. Everybody may place a child into the sandbox.
6. Everybody may move a child to another location in the playground.

Rules #1 and #4-6 are simple rules that can be expressed by setting permissions for read and/or write access for the respective fields. Rules #2 and #3 however are complex and must be handled by defining sub-queries. Figure 6 shows the ELA for the playground.

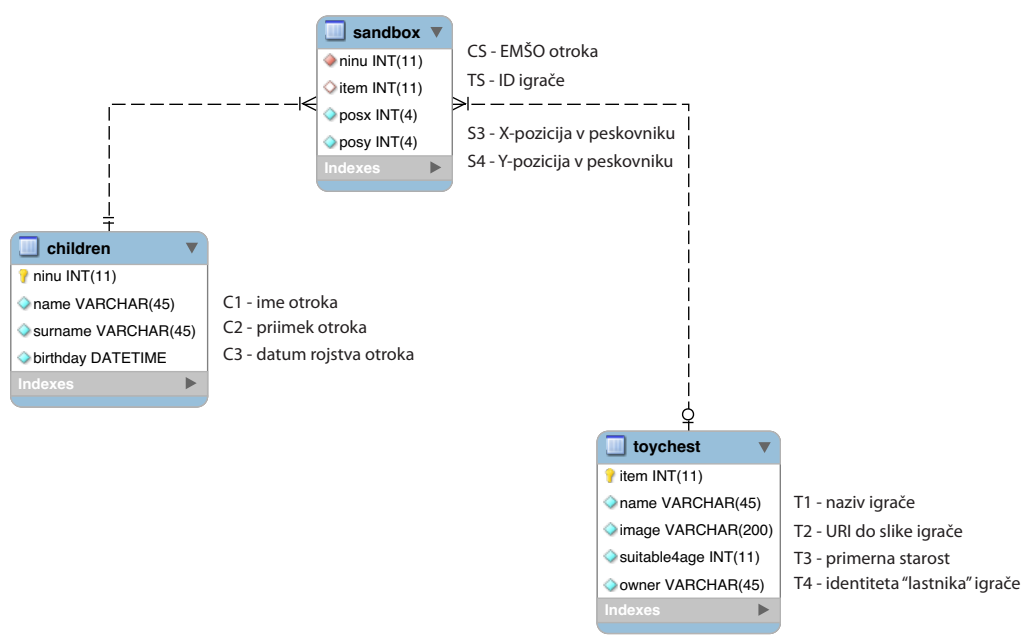

**Figure 2: ER diagram of the *playground* scheme**

Rules #4 & #5 for example would be realized by giving the user "anon" *INSERT* permission to the fields *toychest.{name, image, suitable4age}* (the field *toychest.item* increments automatically), and the same to the fields *sandbox.{ninu, item, posx, posy}*.

The rules #2 and #3 can be solved using the restrictions *#toyInUse* and *#suitableAge*, as shown bolded in Figure 6. Table 2, Table 3 and Table 5 demonstrate slightly simplified how these rules would be applied to incoming requests.

Especially complex is rule #3, which requires that children can play only with toys that are fit for their age; each toy namely has an age restriction and children who are younger than the required age must not use it. However a child that is too young to play with a toy today might be old enough to play with it tomorrow – the rule therefore must be aware of the temporal context. Furthermore, rule #1 restricts anybody from knowing the age of the children, as it mandates the birth dates to be protected; it is hence impossible to know in advance if the child will successfully receive the toy we want to give it.

Nonetheless, the SecSS must enforce all rules. In order to cope with rule #3, the sub-query from the restriction first ad-hoc calculates the current age of the selected child using available functions for time calculations provided by the MySQL RDBMS, as shown in the SQL statement below.

- `SELECT YEAR(FROM_DAYS(DATEDIFF(NOW(), DATE(`**`@bday`**`))))`

Using this technique, the age of a child can be correctly taken into account without revealing it to the requester.

Table 5 shows a further complicated example where a toy is given to a child without the user knowing the relevant data needed to make the desired change. In that case, the information to be updated is calculated at run-time and although only the IDs of the child and toy are provided (instead of the names of each, as required by that simplified example), the update performs correctly.

### The user interface

An important stakeholder group of ss-Gov are service-providers, who facilitate the interaction with ss-Gov registries by providing graphical user interfaces and/or other means of user-friendly interaction (Paulin 2012, 2.2.3). An instantiation of such service in form of an Ajax web application providing a graphical and textual user interface has been developed complementary to the SecSS prototype.

The textual user interface (shown in Figure 8) allows users to design and submit arbitrary SQL queries to the SecSS, and visualizes the responded results in tables. The design for this user interface has been inspired by modern visual tools for interaction with database management systems, like e.g. the MySQL Workbench.

A more advanced way of interaction is possible trough the graphical user interface (Figure 3), which provides an intuitive drag-and-drop experience. In this mode, the sandbox, children and individual toys are visualized and the GUI allows children to be dropped into, dragged out of, or moved inside the sandbox;

also the toys can be dragged onto children. After each action, the corresponding SQL is generated and presented to the user for signing. The signed SQL is packed in the appropriate format and sent to the SecSS; in case of a negative response, the performed action is reversed.

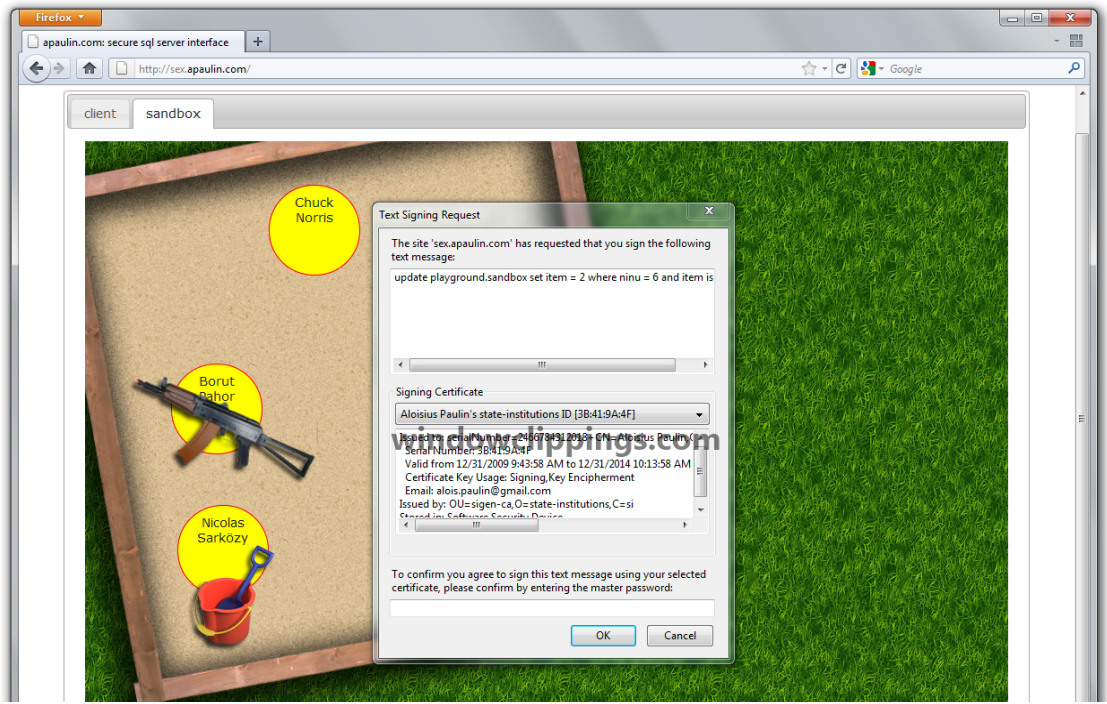

Figure 3: Graphical user interface for manipulating the playground

The communication between the Ajax client and the SecSS uses the formats as described in chapter 3. Each SQL query created directly or indirectly by the user is presented for signing using the built-in functionality provided by the Firefox web browser. Firefox presents the user an intuitive *"what you see is what you sign"* dialog, which allows the user to view the SQL prior to signing it; the user then chooses the desired digital certificate and unlocks the corresponding private key using its password.

The demonstrated client application proves that it is possible to interact with ss-Gov registries in a user-friendly way without compromising the requirements imposed by the ss-Gov governance model.

## 5   Conclusion

In the present article we demonstrated a prototype instance for a ss-Gov registry that complies with the six ss-Gov requirements as defined in (Paulin 2012). We proposed that data based on which eligibilities are calculated and governance is performed can be effectively stored in relational databases to which self-service read and write access is regulated using relational algebra and the SQL standard.

SS-Gov further defines several stakeholders that impose needs that must be fulfilled by any ss-Gov registry instance. The demonstrated SecSS prototype, the complementary prototype visual client application and the ELA document format demonstrate how the stakeholder requirements can be satisfied using modern technology. We evaluated the presented technologies using a scenario of regulating legal relations in a children's playground. We defined six real-world rules that apply to this scenario and successfully demonstrated how they can be enforced using SecSS.

## 6   References

Akram, H., and M. Hoffmann. 2008. "Supports for Identity Management in Ambient Environments - The Hydra Approach." In *3rd International Conference on Systems and Networks Communications, 2008. ICSNC '08*, 371-377. IEEE. doi:10.1109/ICSNC.2008.77.

Angles, Renzo, and Claudio Gutierrez. 2005. "Querying RDF Data from a Graph Database Perspective." In *The Semantic Web: Research and Applications*, ed. Asunción Gómez-Pérez and Jérôme Euzenat, 3532:346-360. Berlin, Heidelberg:          Springer          Berlin          Heidelberg. http://www.springerlink.com/content/jc29tal7pkqybwxl/.

Anley, C. 2002. "Advanced SQL injection in SQL server applications." *White paper, Next Generation Security Software Ltd*.

Anon. 2005. "JavaScript Crypto." *Mozilla Develper Networ*.

Bain, Tony. 2009. "Is the Relational Database Doomed?, 2009." *ReadWriteWeb*. http://www.readwriteweb.com/enterprise/2009/02/is-the-relational-database-doomed.php.

Beckett, David, and Tim Berners-Lee. 2008. "Turtle - Terse RDF Triple Language." *W3C*.          http://www.w3.org/TeamSubmission/2008/SUBM-turtle-20080114/.

Berners-Lee, Tim. 1998. "Relational Databases on the Semantic Web." *DesignIssues*. http://www.w3.org/DesignIssues/RDB-RDF.html.

———. 2009. "Linked Data." *DesignIssues*. http://www.w3.org/DesignIssues/LinkedData.html.

Berners-Lee, Tim, Dan Connolly, Lalana Kagal, Yosi Scharf, and Jim Hendler. 2008. "N3logic: A logical framework for the world wide web." *Theory Pract. Log. Program*. 8 (3) (May): 249–269. doi:10.1017/S1471068407003213.

Berners-Lee, Tim, J. Hendler, O. Lassila, and others. 2001. "The semantic web." *Scientific American* 284 (5): 34–43.

Blythe, S.E. 2005. "Digital signature law of the United Nations, European Union, United Kingdom and United States: Promotion of growth in E-commerce with enhanced security." *Rich. JL & Tech*. 11: 6–8.

Chamberlin, Donald D., and Raymond F. Boyce. 1974. "SEQUEL: A structured English query language." In *Proceedings of the 1974 ACM SIGFIDET (now SIGMOD) workshop on Data description, access and control*, 249–264. SIGFIDET '74. New York, NY, USA: ACM. doi:10.1145/800296.811515. http://doi.acm.org/10.1145/800296.811515.

Codd, E. F. 1970. "A relational model of data for large shared data banks." *Commun. ACM* 13 (6) (June): 377–387. doi:10.1145/362384.362685.

David, B.M., R. Tonicelli, A. Nascimento, D. Amaral, and L. Peotta. 2011. *Secure Single Sign-On and Web Authentication*. Cryptology ePrint Archive.

Dodson, B., D. Sengupta, D. Boneh, and M.S. Lam. 2010. "Secure, consumer-friendly web authentication and payments with a phone." In *Conference* *on Mobile Computing, Applications, and Services (MobiCASE'10), Santa Clara, CA, USA.*

El Maliki, T., and J. -M Seigneur. 2007. "A Survey of User-centric Identity Management Technologies." In *The International Conference on Emerging Security Information, Systems, and Technologies, 2007. SecureWare 2007*, 12-17. IEEE. doi:10.1109/SECUREWARE.2007.4385303.

Garrison, G.B. 1998. "System and method for restricting access to a data table within a database."

Hayes, Jonathan, and Claudio Gutierrez. 2004. "Bipartite Graphs as Intermediate Model for RDF." In *The Semantic Web – ISWC 2004*, ed. Sheila A. McIlraith, Dimitris Plexousakis, and Frank Harmelen, 3298:47-61. Berlin, Heidelberg: Springer Berlin Heidelberg. http://www.springerlink.com/content/n3c41yvamgqtqr39/.

Hevner, A.R., S.T. March, J. Park, and S. Ram. 2004. "Design Science in Information Systems Research." *Management Information Systems Quarterly* 28 (1): 6.

ISO/IEC 9075-1. 2008. "Information technology - Database design - SQL - Part 1: Framework (SQL/Framework)." ISO.

Lampropoulos, K., and S. Denazis. 2011. "Identity management directions in future internet." *IEEE Communications Magazine* 49 (12) (December): 74-83. doi:10.1109/MCOM.2011.6094009.

Mackey, Alex. 2010. "Windows Communication Foundation." In *Introducing .NET 4.0*, by Alex Mackey, 159-173. Berkeley, CA: Apress. http://www.springerlink.com/content/q3u44t4021732533/.

March, Salvatore T., and Gerald F. Smith. 1995. "Design and natural science research on information technology." *Decis. Support Syst.* 15 (4) (December): 251–266. doi:10.1016/0167-9236(94)00041-2.

Miller, J. 2006. "Yadis specification." *Yadis.* http://yadis.org/.

Miller, S. P, B. C Neuman, J. I Schiller, and J. H Saltzer. 1987. "Kerberos authentication and authorization system." *IN PROJECT ATHENA TECHNICAL PLAN*. http://citeseerx.ist.psu.edu/viewdoc/summary?doi=10.1.1.91.7727.

Patel-Schneider, P.F. 2010. "RDF: Back to the Graph." In  Palo Alto: NCBO.

Paulin, Alois. 2012. "Towards Self-Service Governance by Means of Information Technology." *arXiv:1201.0882* (January 4). http://arxiv.org/abs/1201.0882.

Recordon, David, and Drummond Reed. 2006. "OpenID 2.0: a platform for user-centric identity management." In *Proceedings of the second ACM workshop on Digital identity management*, 11–16. DIM  '06. New York, NY, USA: ACM. doi:10.1145/1179529.1179532. http://doi.acm.org/10.1145/1179529.1179532.

Seaborne, A., G. Manjunath, C. Bizer, J. Breslin, S. Das, I. Davis, S. Harris, et al. 2008. "SPARQL/Update: A language for updating RDF graphs." *W3C Member Submission* 15.

Veljković, N., S. Bogdanović-Dinić, and L. Stoimenov. 2011. "Municipal Open Data Catalogues." In *Proceedings of the International Conference for E-Democracy and Open Government*. Krems: Donau-Universität.

Vicknair, Chad, Michael Macias, Zhendong Zhao, Xiaofei Nan, Yixin Chen, and Dawn Wilkins. 2010. "A comparison of a graph database and a relational database: a data provenance perspective." In *Proceedings of the 48th*

*Annual Southeast Regional Conference*, 42:1–42:6. ACM SE ʼ10. New York, NY, USA: ACM. doi:10.1145/1900008.1900067. http://doi.acm.org/10.1145/1900008.1900067.

Vitvar, Tomáš, Vassilios Peristeras, and Konstantinos Tarabanis. 2010. "Semantic Technologies for E-Government: An Overview." In *Semantic Technologies for E-Government*, ed. Tomas Vitvar, Vassilios Peristeras, and Konstantinos Tarabanis, 1-22. Berlin, Heidelberg: Springer Berlin Heidelberg. http://www.springerlink.com/content/x6268p5t52444747.

Wieringa, Roel. 2009. "Design science as nested problem solving." In *Proceedings of the 4th International Conference on Design Science Research in Information Systems and Technology*, 8:1–8:12. DESRIST ʼ09. New York, NY, USA: ACM. doi:10.1145/1555619.1555630. http://doi.acm.org/10.1145/1555619.1555630.

**Figure 4: Electronic rules inside the element `Restriction` of an ELA**

**Figure 5: Managing read/write access to the field `sandbox.toy`**

**Figure 6: Rules for the "Playground" scenario stored in the ELA**

```xml
<?xml version="1.0" encoding="utf-8" ?>
<Configuration>
    <Connections></Connection>
    <Restrictions>
        <Restriction
            Id="toyInUse" type="INSERT/UPDATE"
            table="sandbox" field="@item" use="NOT IN">
            <var field="item" name="@item" />
            <sql>
            <![CDATA[
                    SELECT s.item FROM playground.sandbox s
            ]]>
            </sql>
            <justification>
                    A toy that is in use by a child must not be
                    given to another child.
            </justification>
        </Restriction>
        <Restriction
            Id="suitableAge" type="INSERT/UPDATE"
            table="sandbox" field="@item" use="IN">
            <var field="ninu" name="@ninu" />
            <var field="item" name="@item" />
            <sql>
            <![CDATA[
                    SELECT t.item FROM playground.toychest t
                    WHERE t.suitable4age <=
                        (SELECT YEAR(FROM_DAYS(
                          DATEDIFF(NOW(), DATE(c.birthday))))
                         FROM playground.children c
                         WHERE c.ninu = @ninu)
            ]]>
            </sql>
            <justification>
                    A child can only play with toys for which
                    it is old enough.
            </justification>
        </Restriction>
    </Restrictions>
    <Permissions>
        <Schema name="playground">
          <Table name="sandbox">
            <Field name="ninu">
               <Permission user="anon" type="INSERT" />
               <Permission user="anon" type="SELECT" />
            </Field>
            <Field name="item">
               <Permission user="anon" type="SELECT" />
               <Permission user="anon" type="INSERT">
                  <Apply-Restriction ref="#suitableAge" />
               </Permission>
               <Permission user="anon" type="UPDATE">
                  <Apply-Restriction ref="#suitableAge" />
               </Permission>
            </Field>
            <Field name="posx">
               <Permission user="anon" type="SELECT" />
```

```
                <Permission user="anon" type="INSERT" />
                <Permission user="anon" type="UPDATE" />
            </Field>
            <Field name="posy">
                <Permission user="anon" type="SELECT" />
                <Permission user="anon" type="INSERT" />
                <Permission user="anon" type="UPDATE" />
            </Field>
        </Table>
        <Table name="children">
            <Field name="ninu">
                <Permission user="anon" type="SELECT" />
            </Field>
            <Field name="name">
                <Permission user="anon" type="SELECT" />
            </Field>
            <Field name="surname">
                <Permission user="anon" type="SELECT" />
            </Field>
        </Table>
        <Table name="toychest">
            <Field name="*">
                <Permission user="@owner" type="DELETE" >
                <!-- the "@" sign before the field name indicates
                    that the identity is stored in the
                    corresponding field -->
                <justification>
                    The "owner" of a toy may remove the toy from
                    the toy-chest. The database must feature a
                    correctly set FK-constraint to prevent removing
                    a toy that is in use.
                </justification>
                </Permission>
                <Permission user="###(admin)###" type="UPDATE" />
            </Field>
            <Field name="item">
                <Permission user="anon" type="SELECT" />
            </Field>
            <Field name="name">
                <Permission user="anon" type="SELECT" />
                <Permission user="anon" type="INSERT" />
            </Field>
            <Field name="image">
                <Permission user="anon" type="SELECT" />
                <Permission user="anon" type="INSERT" />
            </Field>
            <Field name="suitable4age">
                <Permission user="anon" type="SELECT" />
                <Permission user="anon" type="INSERT" />
            </Field>
        </Table>
      </Schema>
   </Permissions>
</Configuration>
```

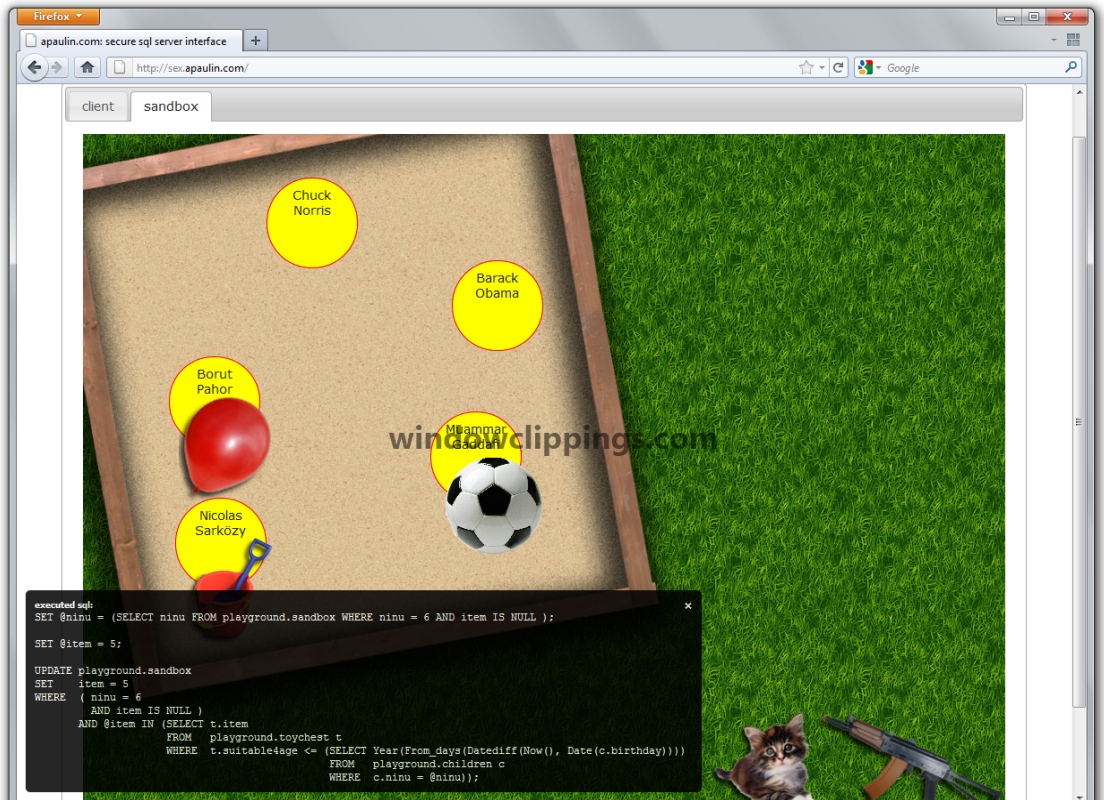

**Figure 7: The SQL as it has been transformed and executed is presented after a successful transaction.**

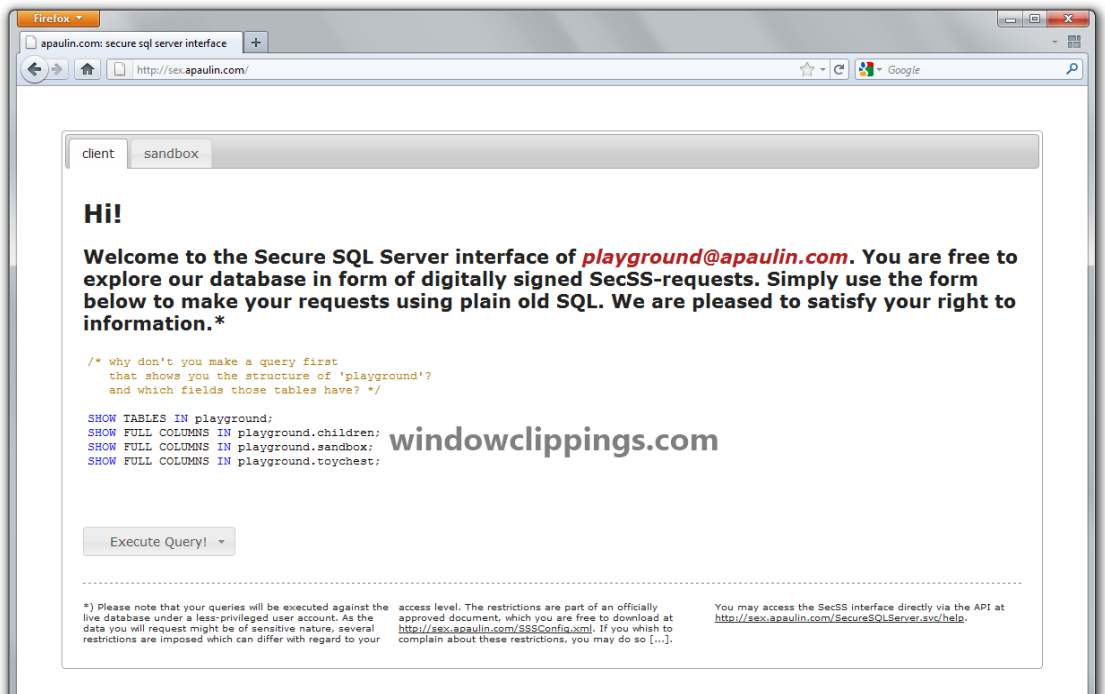

**Figure 8: The SecSS prototype client allows also text-based access to data**